\documentclass{article}
\usepackage{spconf,amsmath,graphicx,hyperref}
\usepackage{graphicx}
\usepackage{amsmath}
\usepackage{amssymb}
\usepackage{subcaption}
\usepackage{multirow}
\usepackage{booktabs}
\usepackage{xcolor}
\usepackage{float}
\usepackage{tabularx}
\usepackage{makecell} 
\usepackage{amsmath}    
\usepackage{amssymb}    
\usepackage{bm}         

\makeatletter
\renewcommand{\name}[1]{%
  \gdef\@name{%
    \em 
    \begingroup
    \parbox[t]{0.95\textwidth}{\centering #1}%
    \endgroup
  }%
}
\makeatother

\title{MECap-R1: Emotion-aware Policy with Reinforcement Learning for Multimodal Emotion Captioning}

\name{Haoqin Sun\textsuperscript{1,2}, Chenyang Lyu\textsuperscript{2,*}, Xiangyu Kong\textsuperscript{3}, Shiwan Zhao\textsuperscript{1}, Jiaming Zhou\textsuperscript{1}, Hui Wang\textsuperscript{1}, Aobo Kong\textsuperscript{1},  Jinghua Zhao\textsuperscript{1},   Longyue Wang\textsuperscript{2}, Weihua Luo\textsuperscript{2}, Kaifu Zhang\textsuperscript{2}, Yong Qin\textsuperscript{1,*}\thanks{This work has been supported by the National Key R\&D Program of China through (Grant No.2022ZD0116307) and NSF China (Grant No.62271270). Email: sunhaoqin@mail.nankai.edu.cn.}\vspace{15pt}}

\vspace{15pt}

\address{$^{1}$TMCC, College of Computer Science, Nankai University, Tianjin, China \\ $^{2}$Alibaba International Digital Commerce \\ $^{3}$University of Exeter, Exeter, United Kingdom}
\begin{document}

\maketitle

\begin{abstract}
Speech Emotion Captioning (SEC) has emerged as a notable research direction. The inherent complexity of emotional content in human speech makes it challenging for traditional discrete classification methods to provide an adequate representation. Consequently, utilizing natural language to describe speech emotions presents a novel avenue for more effectively capturing and expressing affect. In this paper, we propose MECap-R1, a pioneering emotion-aware policy with reinforcement learning for multimodal emotion captioning. By employing Group Relative Policy Optimization with emotion-aware reward (Emo-GRPO), the framework precisely captures the emotion and semantic features, thereby addressing the shortcomings of rigid rules in handling the dynamic and flexible nature of captions. Experimental results on the EmotionTalk dataset demonstrate that MECap-R1 performs well in generating emotion descriptions and achieves substantial gains in both accuracy and diversity.
\end{abstract}

\section{Introduction}
\label{sec:intro}
Speech Emotion Recognition (SER)~\cite{sun2024iterative, sun2025enhancing} aims to classify emotions into predefined discrete labels based on speech acoustics. While SER has advanced, its inherent limitation of simplifying emotions into discrete categories remains a significant challenge. This discretization fails to capture emotional nuances, limiting performance with ambiguous emotions and resulting in the loss of valuable semantic information. To address this, the paradigm is shifting from emotion classification to emotion captioning generation~\cite{zhu2024unistyle,sun25c_interspeech, chandra25_interspeech}. The latter task generates descriptive text that naturally and richly conveys emotion. Compared to classification, emotion captioning offers more comprehensive and detailed emotional insights, providing a path toward human-computer interaction systems with greater empathy and understanding.

In recent years, a number of works have drawn inspiration from the task paradigm of automated audio captioning~\cite{tian25_interspeech, deshmukh2024training} to achieve more nuanced emotion descriptions. For example, Xu et al.~\cite{xu2024secap} propose the SECap framework, which pioneered the use of natural language captions to describe emotions in speech. At its core, the technique leverages mutual information learning to decouple content and emotional features. Liang et al.~\cite{liang2024aligncap} propose AlignCap, a method that ensures high-quality captions through the integration of KD-Regularization and PO-Regularization. Additionally, Chen et al.~\cite{chen25i_interspeech} introduce a LLaMA-based~\cite{dubey2024llama} emotion captioning framework to explore the underlying connections between SEC and SER. Despite recent progress, current research has two key limitations. First, models fail to fully use visual information. Second, methods using reinforcement learning, like AlignCap, lack effective reward mechanisms, which limits the authenticity and diversity of the generated content.

Following the promising results of DeepSeek-R1~\cite{guo2025deepseek}, we explore the potential of rule-based reinforcement learning for improving the perception of emotions. In this paper, we propose MECap-R1, a pioneering multimodal emotional captioning framework with Emo-GRPO. For the first stage, we use supervised fine-tuning (SFT) for model cold-starting. Given the diversity and flexibility of emotion descriptions, hard rules are not applicable. Therefore, in the second stage, we adopt reinforcement learning (GRPO) and tailor a unique emotion-aware reward mechanism for the post-training process. This mechanism works by creating an emotion anchor space to accurately capture the emotional and semantic features of the text. The experimental results on the EmotionTalk dataset demonstrate that the MECap-R1 model achieves greater accuracy and diversity in emotion descriptions.

To the best of our knowledge, this study marks the first to systematic application of GRPO algorithm and visual information for SEC task. We expect this work to provide a new perspective for reinforcement learning-based approaches to speech emotion understanding.

\begin{figure*}[t]
  \centering
  \includegraphics[width=6.1in]{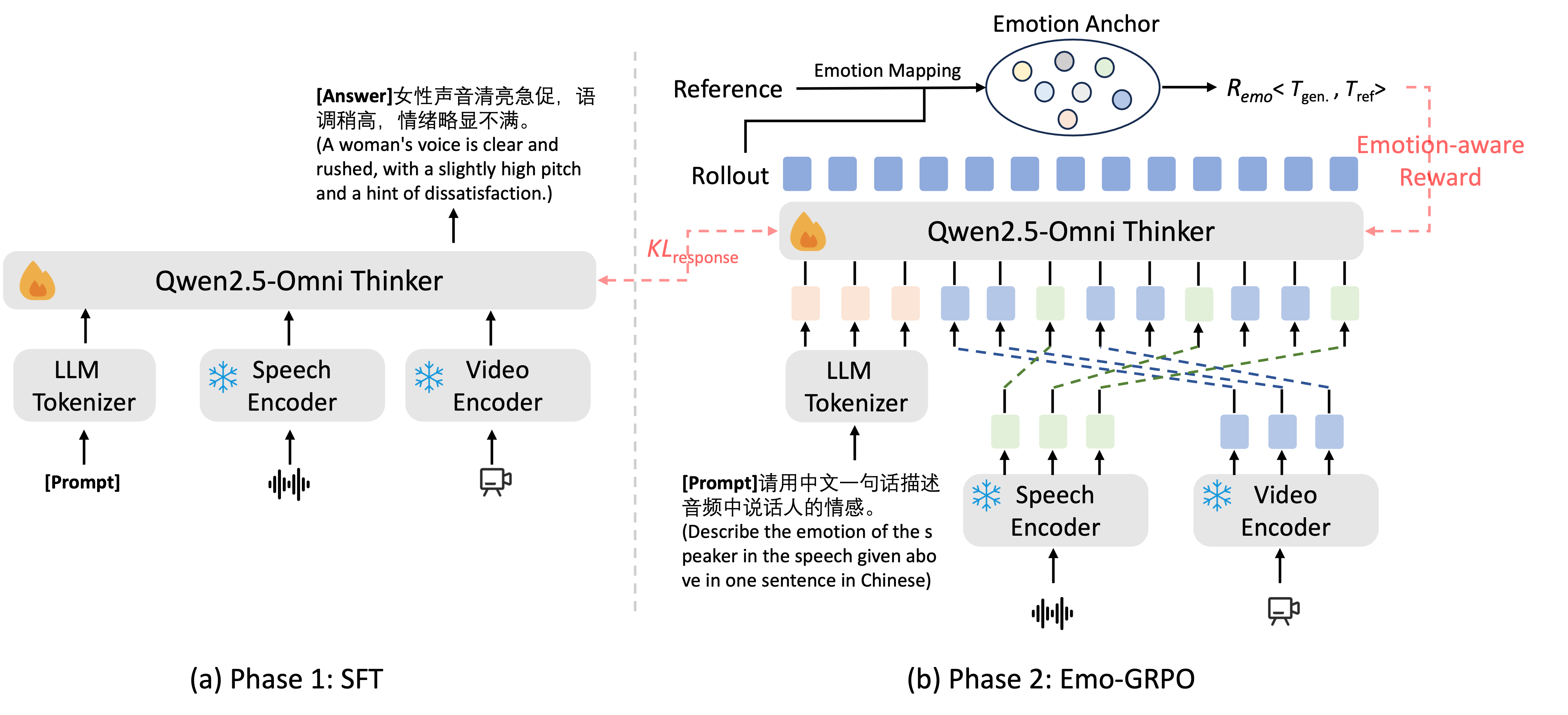}
  \caption{The overall structure of the proposed framework.}
  \label{fig:framework}
\end{figure*}

\section{MECap-R1}
\label{sec:method}
Our proposed model, MECap-R1, is trained in two principal phases. First, we establish a foundational model through supervised learning. Second, we refine its policy using a reward function specifically engineered to capture nuanced emotional semantics.

\subsection{Phase 1 - Cold-Start: Supervised Fine-Tuning (SFT)}
In the initial phase, we establish a foundational captioning capability for the generative model, denoted as $\mathcal{G}_{\theta}$ with parameters $\theta$. This is achieved by supervised fine-tuning (SFT), which corresponds to optimizing the model's parameters via Maximum Likelihood Estimation (MLE).

Given a dataset of $N$ training samples, where each sample $i$ consists of a multimodal context $M_i$ and a target text sequence $Y_i = (y_{i,1}, \dots, y_{i,T_i})$, the objective is to find the parameters $\theta^*$ that maximize the conditional log-likelihood of the target sequences. This is formulated as the minimization of the negative log-likelihood loss, $\mathcal{L}_{\text{SFT}}$:
\begin{equation}
\mathcal{L}_{\text{SFT}}(\theta) = - \sum_{i=1}^{N} \log P(Y_i | M_i; \theta).
\end{equation}
Assuming an autoregressive framework, this objective decomposes into the sum of log-probabilities for each token:
\begin{equation}
\mathcal{L}_{\text{SFT}}(\theta) = - \sum_{i=1}^{N} \sum_{t=1}^{T_i} \log P(y_{i,t} | y_{i,<t}, M_i; \theta).
\end{equation}
This phase equips the model $\mathcal{G}_{\theta}$ with a strong prior over the distribution of plausible descriptive text. To foster the generation of more diverse and emotionally salient captions, we introduce a subsequent refinement stage based on reinforcement learning.

\subsection{Phase 2 Emo-GRPO: Emotion-Aware Reward Reinforcement Modeling}
To guide the model beyond simple likelihood maximization, we employ the GRPO reinforcement learning algorithm, which refines the model's policy using a meticulously crafted reward function. We structure this reward by first defining a specialized semantic subspace sensitive to emotional concepts.

Let $\mathcal{T}$ be the space of all text sequences and let $\mathcal{E}: \mathcal{T} \to \mathbb{R}^D$ be a semantic embedding function that maps any given text into a $D$-dimensional latent semantic space. This function is realized by a pre-trained Sentence-BERT encoder.\footnote{\url{https://huggingface.co/shibing624/text2vec-base-chinese}}

\subsubsection{Emotion Anchor Construction}
For a predefined set of $n$ discrete emotion categories $\{E_i\}_{i=1}^n$, we curate a corresponding set of lexical corpora $\{W_i\}_{i=1}^n$, where each $W_i \subset \mathcal{T}$ contains vocabulary terms semantically indicative of emotion $E_i$.

We then define an \textbf{Emotion Anchor}, $\bm{a}_i \in \mathbb{R}^D$, for each category $E_i$. This anchor is formulated as the \textbf{centroid} of the embedded lexical set within the latent space, effectively capturing the central semantic concept of that emotion:
\begin{equation}
\bm{a}_i = \frac{1}{|W_i|} \sum_{w \in W_i} \mathcal{E}(w).
\end{equation}
The collection of these anchors, $\{\bm{a}_i\}_{i=1}^n$, forms a conceptual basis within $\mathbb{R}^D$ that represents the principal axes of our target emotion space.

\subsubsection{Emotion Coordinate Mapping}
We introduce a mapping function $\Phi: \mathbb{R}^D \to \mathbb{R}^n$ that projects any semantic vector from the general $D$-dimensional space into our newly defined $n$-dimensional \textbf{Emotion Coordinate Space}. 

For any text $T$, its representation in this space, denoted by the vector $\bm{c}_T \in \mathbb{R}^n$, is obtained by first embedding it into the latent space to get $\bm{t} = \mathcal{E}(T)$, and then applying the projection map $\Phi$:
\begin{equation}
\bm{c}_T = \Phi(\bm{t}).
\end{equation}
The $i$-th component of this projected vector, $(\bm{c}_T)_i$, is defined as the normalized projection of $\bm{t}$ onto the corresponding emotion anchor $\bm{a}_i$, calculated via cosine similarity:
\begin{equation}
(\bm{c}_T)_i = \frac{\bm{t} \cdot \bm{a}_i}{\|\bm{t}\| \|\bm{a}_i\|}.
\end{equation}
This procedure transforms any text into a vector of emotional intensities. To formulate our reward, we compare the generated caption, $T_{\text{gen}}$, with the ground-truth reference, $T_{\text{ref}}$. The emotion-aware reward, $R_{\text{emo}}$, is defined as the cosine similarity of their respective representations within the Emotion Coordinate Space:
\begin{equation}
R_{\text{emo}}(T_{\text{gen}}, T_{\text{ref}}) = \frac{\Phi(\mathcal{E}(T_{\text{gen}})) \cdot \Phi(\mathcal{E}(T_{\text{ref}}))}{\|\Phi(\mathcal{E}(T_{\text{gen}}))\| \|\Phi(\mathcal{E}(T_{\text{ref}}))\|}.
\end{equation}
This reward measures the structural alignment of the emotional content between the two texts, rather than their surface-level lexical overlap.

\subsubsection{Final Reward Objective}
To ensure both semantic fidelity and linguistic quality, we construct a composite reward signal. This signal linearly combines our emotion-aware reward with established metrics for textual quality. Specifically, we incorporate BLEU ($S_{\text{BLEU}}$) for n-gram precision and SPICE ($S_{\text{SPICE}}$) for semantic propositional content.

The total reward function, $R_{\text{total}}$, which provides the learning signal for the GRPO algorithm, is a weighted amalgamation of these components:
\begin{equation}
R_{\text{total}} = \alpha R_{\text{emo}} + \beta \left( S_{\text{BLEU}} + S_{\text{SPICE}} \right),
\end{equation}
where $\alpha, \beta \in \mathbb{R}^+$ are hyperparameters that control the relative importance of conforming to the emotional structure versus maintaining grammatical and semantic correctness. This multi-objective function guides the model to produce captions that are not only accurate but also emotionally resonant and structurally sophisticated.

\section{Experiments}
\label{sec:pagestyle}
\subsection{Dataset}
Developed by Nankai University, EmotionTalk~\cite{sun2025emotiontalk} is an interactive Chinese multimodal dataset designed specifically for emotion analysis. The dataset compiles dialogues from 19 professional actors, totaling 23.6 hours and including 19,250 utterances across three modalities: audio, video, and text. It features a rich annotation system, including four types of emotional speaking style captions.

\subsection{Experimental Setup}

\textbf{Training Settings:} during the SFT stage, we use the AdamW optimizer with a learning rate of 1e-4, a batch size of 1, and a gradient accumulation step of 2. The LoRA rank is set to 8. In the emo-GRPO stage, the batch size remains 1, while gradient accumulation is increased to 4, and a warm-up ratio of 0.05 is applied. The learning rate, optimizer, and LoRA configurations are consistent with those in the SFT stage. The remaining hyperparameter settings are as shown in Table~\ref{tab:settings}.

\textbf{Baseline:} we select BART~\cite{lewis2019bart}, GPT-2~\cite{lagler2013gpt2}, and Qwen-2~\cite{team2024qwen2} as baseline models, which use HuBERT as encoder.

\begin{table}[t]
\caption{Hyperparameter Settings}
\centering
\label{tab:settings}
\begin{tabular}{cc}
\toprule
Gradient Accumulation & 4 \\
Rollout & 4 \\
KL Coefficient & 0.5 \\
Max Response Length & 2048 \\
Temperature & 1.0 \\
\bottomrule
\end{tabular}
\end{table}

\begin{figure}[t]
  \centering
  \includegraphics[width=3.0in]{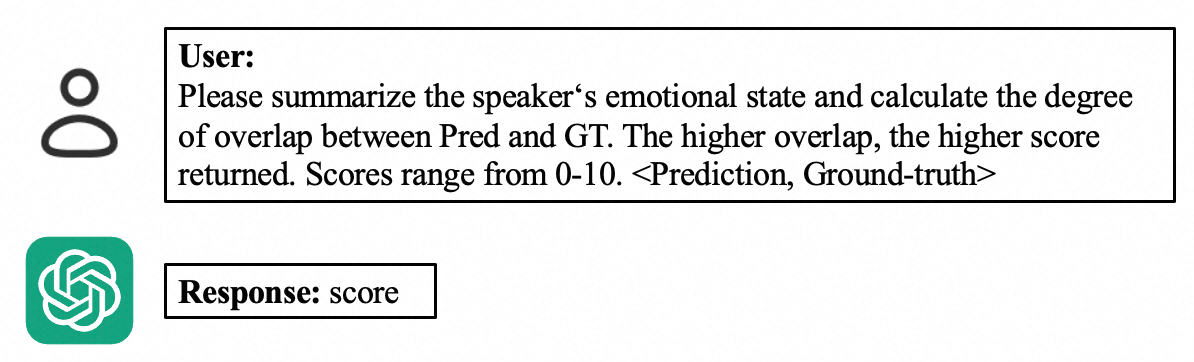}
  \caption{Prompt for GPT-4 Evaluation.}
  \label{fig:evaluate}
\end{figure}

\textbf{Evaluation Metrics:}
In this study, we align with the AAC task by employing objective evaluation metrics, including BLEU, ROUGE$_l$, METEOR, SPIDER, and Vocab. Additionally, as shown in Fig.~\ref{fig:evaluate}, we use GPT-4~\cite{achiam2023gpt} to assess the relevance and summarization of the emotional captions.

\begin{table*}[t]
\centering
\caption{Comparison of emotion captioning performance. Vocab is the unique vocabulary of captions. * denotes zero-shot.}
\begin{tabular}{lcccccccccc}
\toprule
\textbf{Model} & \textbf{BLEU$_1$} & \textbf{BLEU$_2$} & \textbf{BLEU$_3$} & \textbf{BLEU$_4$} & \textbf{ROUGE$_l$} & \textbf{METEOR} & \textbf{SPIDER} & \textbf{Vocab} \\ 
\midrule
BART~\cite{lewis2019bart} & - & - & - & 1.8 & 46.9 & 23.3  & \textbf{23.0} & -\\
GPT-2~\cite{lagler2013gpt2} & - & - & - & 1.5 & 46.2 & 21.4 & 22.7 & -\\
Qwen-2~\cite{team2024qwen2} & - & - & - & 3.3 & 53.5& 26.8  & 12.1 & -\\
Qwen2.5-Omni$^*$~\cite{xu2025qwen2} & 26.2 & 0.9 & 0.3 & 0.0 & 36.1 & 13.1 & 10.8 & 100 \\
\textbf{Mecap-R1} & \textbf{54.6} & \textbf{27.0} & \textbf{18.1} & \textbf{7.2} & \textbf{54.7} & \textbf{29.3} & 12.8 & 229 \\ \hline
w/o SFT & 53.5 & 24.2 & 15.8 & 0.0 & 53.7 & 22.0 & 11.6 & \textbf{557} \\
w/o emo-GRPO & 49.9 & 18.3 & 11.1 & 3.5 & 50.5 & 19.5 & 12.7 & 181\\
w/o emotion-aware reward & \textbf{54.6} & 26.6 & 17.7 & 6.6 & 54.6 & 29.1 & 12.7 & 209\\
w/o video & 53.1 & 25.5 & 16.6 & 5.4 & 53.2 & 27.7 & 11.3 & 209\\
\bottomrule
\end{tabular}
\label{tab:res}
\end{table*}

\begin{table*}[h!]
    \centering
    \caption{Automatic emotion evaluation cases. The overall emotion score is 5.43 for S1 and \textbf{5.84} for S2.} 
    \label{tab:emo}
    \begin{tabularx}{\textwidth}{@{}l XX@{}}
        \toprule
        & \textbf{G00006\_35\_12\_041.wav} & \textbf{G00001\_12\_01\_030.wav} \\
        \midrule
        S1: MECap-R1 w/o emotion-aware reward & The man's voice is loud and clear, spoken at a relatively fast pace, with heightened emotional excitement. (GPT-4 Score: 4.0)&The man's voice is bright and penetrating, spoken at a relatively fast pace, with an upward inflection indicating urgency. (GPT-4 Score: 3.0)\\
        \midrule
        S2: MECap-R1 & The male's voice is loud and clear, the pace is quick, the tone rises, and the emotion is agitated with a forceful tone. (GPT-4 Score: 5.0)& The male's voice is clear and magnetic, the pace is moderate, and the tone is calm and natural. (GPT-4 Score: 8.0)\\
        \midrule
        Ground-truth & The adult male's tone is rising, his voice is high-pitched, and the slightly fast pace suggests dissatisfaction.& The male's voice is magnetic and steady, the pace is moderate, showing a relaxed and calm emotion. \\
        \bottomrule
    \end{tabularx}
\end{table*}

\subsection{Main Results}
Table~\ref{tab:res} presents the experimental results for the baseline and our proposed method.

\textbf{Comparison experiment:} the top half of Table~\ref{tab:res} shows that MECap-R1 significantly outperforms other models on BLEU, ROUGE$_L$, and METEOR metrics, indicating a clear advantage in text generation accuracy and semantic quality. Although its SPIDER score is slightly lower than the BART and GPT-2 models, its larger vocabulary allows for richer language expression. Overall, MECap-R1 effectively enhances the quality and diversity of generated emotion captions.

\textbf{Ablation study:} to validate the effectiveness of each component in the MECap-R1 model, we conduct in-depth ablation studies. The removal of SFT (w/o SFT) causes the BLEU-4 metric to drop to 0.0, which indicates that SFT is foundational for the model's cold-start phase. Similarly, all metrics decrease when the video modality (w/o video) is removed, proving that video provides essential context as a supplementary source of information. 

Most importantly, removing emo-GRPO leads to a significant drop in all metrics, with words diversity also decreasing from 229 to 181. This powerfully demonstrates that emo-GRPO is a core component for improving description quality and diversity, as it enables the model to learn to generate more nuanced and emotionally profound descriptions by comparing multiple response sets. 

Furthermore, when removing the emotion-aware reward, we use GPT-4 to evaluate emotion overlap (see Table~\ref{tab:emo}). The results show that not only do our objective metrics drop slightly, but the overall emotion overlap of the descriptions also decreases. For instance, in the first case , S1 only provides a general description, whereas S2 accurately captures nuances like "tone rises" and "agitated emotion," resulting in a higher emotion score. The second case is even more striking: the description generated by the S1 model directly contradicts the actual emotion (“calm”), resulting in an extremely low emotion score. This confirms the effectiveness of the emotion-aware reward.

\vspace{-3mm}

\section{Conclusion}
\label{sec:con} 
To overcome the limitations of rule-based reinforcement learning in generating emotion captions, we propose a pioneering framework, MECap-R1. Its key innovation lies in emo-GRPO, a novel emotion-aware policy with reinforcement learning, which is achieved by calculating the similarity between the generated captions and the ground-truth within the emotional embedding space. Experiments on the EmotionTalk dataset demonstrate that MECap-R1 achieves ideal performance in caption diversity and emotion matching. Overall, this study explored the potential of rule-based reinforcement learning for emotion-aware policy learning. We hope this work provides new insights for the application of reinforcement learning in the affective computing.

\clearpage
\bibliographystyle{IEEEbib}
\bibliography{strings,refs}

\end{document}